\documentclass[10pt,letterpaper,floatfix,aps,pra,reprint,superscriptaddress,amsmath,amsfonts,longbibliography]{revtex4-1}

\usepackage{graphicx}
\usepackage{braket}

\begin{document}
	
\title{Linear feedback stabilization of a dispersively monitored qubit}

\author{Taylor Lee Patti}
\affiliation{Schmid College of Science and Technology, Chapman University, Orange, CA 92866, USA}
\affiliation{Institute for Quantum Studies, Chapman University, Orange, CA 92866, USA}
\affiliation{Department of Physics, Harvard University, Cambridge, Massachusetts 02138, USA}
\author{Areeya Chantasri}
\affiliation{Department of Physics and Astronomy, University of Rochester, Rochester, New York 14627, USA}
\affiliation{Center for Coherence and Quantum Optics, University of Rochester, Rochester, New York 14627, USA}
\affiliation{Centre for Quantum Dynamics, Griffith University, Nathan, Queensland 4111, Australia}
\author{Luis Pedro Garc\'ia-Pintos}
\affiliation{Schmid College of Science and Technology, Chapman University, Orange, CA 92866, USA}
\affiliation{Institute for Quantum Studies, Chapman University, Orange, CA 92866, USA}
\author{Andrew N. Jordan}
\affiliation{Institute for Quantum Studies, Chapman University, Orange, CA 92866, USA}
\affiliation{Department of Physics and Astronomy, University of Rochester, Rochester, New York 14627, USA}
\affiliation{Center for Coherence and Quantum Optics, University of Rochester, Rochester, New York 14627, USA}
\author{Justin Dressel}
\affiliation{Schmid College of Science and Technology, Chapman University, Orange, CA 92866, USA}
\affiliation{Institute for Quantum Studies, Chapman University, Orange, CA 92866, USA}

\begin{abstract}
The state of a continuously monitored qubit evolves stochastically, exhibiting competition between coherent Hamiltonian dynamics and diffusive partial collapse dynamics that follow the measurement record. We couple these distinct types of dynamics together by linearly feeding the collected record for dispersive energy measurements directly back into a coherent Rabi drive amplitude. Such feedback turns the competition cooperative, and effectively stabilizes the qubit state near a target state. We derive the conditions for obtaining such dispersive state stabilization and verify the stabilization conditions numerically. We include common experimental nonidealities, such as energy decay, environmental dephasing, detector efficiency, and feedback delay, and show that the feedback delay has the most significant negative effect on the feedback protocol. Setting the measurement collapse timescale to be long compared to the feedback delay yields the best stabilization.
\end{abstract}

\maketitle

\section{Introduction}

Modern efforts to build a quantum computer have recently enabled the time-continuous measurement of quantum state trajectories \cite{BookGardiner2,BookCarmichael,BookWiseman,Jacobs2006} using superconducting circuit quantum-electrodynamics (cQED) \cite{Gambetta2008,Korotkov2014,Korotkov2016}. For quantum non-demolition (QND) measurements, such as dispersive transmon energy level measurements \cite{koch07}, a time-continuous noisy signal gradually reveals information over time about the stationary eigenstates of the measurement. Learning this information causes a gradual collapse of the qubit state over time to one of the eigenstates of the measurement, which can be observed \cite{Kater2013,Roch2014,ibarcq2015} by processing the noisy measurement record. Notably, such gradual collapse is not monotonic, but rather stochastic, with some temporal increments effectively erasing the information gathered by previous increments and ``uncollapsing'' the state \cite{Korotkov2006,Korotkovexp2008,Jordan2010,Schindler2013}. Nevertheless, the random walk of the qubit state does eventually fully collapse to one of the measurement eigenstates, where it remains. Continuing to monitor an eigenstate after collapse helps protect it from environmental decay due to the quantum Zeno effect \cite{Misra1977,Facchi2008,Slichter2016}, which suppresses other coherent evolution via continual recollapse.

The stochastic collapse evolution from continuous monitoring generally competes with other coherent Hamiltonian evolution, which produces complicated state dynamics that have rich statistical correlations \cite{Weber2014,Chantasri2013,jordan2016fluor,Chantasri2015,Naghiloo2017,Jordan2005,chantasri2016,Lewalle2017}. Similarly, simultaneously monitoring multiple non-commuting observables produces nontrivial competitive dynamics \cite{Ruskov2010,Hacohen-Gourgy2016,Garcia-Pintos-2016,Atalaya2017}. In either case, collapse to a stable eigenstate is prevented. Remarkably, it has been shown that such dynamical competition can be made cooperative by feeding a suitably processed noisy readout back into the controlled Hamiltonian dynamics \cite{Wiseman1993feedback,Wiseman1994,Hofmann1998,Doherty1999,wangwiseman2001,Wiseman2002bayes,Ruskov2002,Ruskov2004,Korotkov2005,Korotkov2005-2,wang2005entfeed,Schirmer2010,Chia2011}. Similar to classical control theory, the random perturbations induced by the stochastic dynamics can be compensated by the adaptive control Hamiltonian to produce customized dynamics that approximate a desired outcome. Unlike classical control theory, such measurement-based quantum feedback control can fundamentally leverage the peculiarities of quantum state collapse in ways that have no classical analog \cite{Doherty2000}. For example, quantum states may be more rapidly purified through a clever application of feedback \cite{Jacobs2003,Combes2006,Wiseman2006,Jordan2006,Combes2008,Ruskov2012}. A variety of quantum feedback control schemes have been implemented within the past six years, including photon number state preparation \cite{Sayrin2011}, continuous superconducting Rabi oscillation stabilization \cite{Korotkov2012}, discrete feedback control of superconducting qubits \cite{Riste2012,Ibarcq2013}, entangled state generation \cite{Riste2013,Shankar2013,Lin2013,Leghtas2015,Schwartz2016}, discrete reversal of continuous random phase drift \cite{deLange2014}, and continuous qubit state stabilization via both dissipative bath engineering \cite{murch12qbe} and feedback from fluorescence measurements \cite{Ibarcq2016}.

In this paper, we theoretically revisit the latter task of state stabilization, but for a transmon qubit undergoing continuous linear feedback of a dispersive energy measurement record in a realistically modeled environment that includes dephasing, energy-decay, measurement inefficiency, signal filtering, and feedback delay. The goal of such a stabilization protocol is to effectively alter the eigenstates of the measurement collapse to target an arbitrary desired state, while preserving the main features of the quantum Zeno effect that protect that state against deterioration from environmental factors. Unlike previously considered dissipative bath engineering \cite{murch12qbe} or fluorescence-based feedback \cite{wangwiseman2001,Ibarcq2016} methods, such dispersive feedback is based on a fundamentally QND measurement. As such, the symmetry of the non-feedback collapse dynamics alters the permissible state stabilization regions dramatically from those produced by the asymmetric backaction of environmental dissipation or energy-decay from fluorescence. The QND nature of the dispersive measurement also makes it natural to compare the differences between ensemble-average state stabilization and individual trajectory stabilization. We thus pose the stabilization problem in two contrasting ways: First, we suppose an experimenter only knows the applied feedback parameters and does not monitor individual measurement records, and so can only prepare a targeted ensemble-averaged state on demand using feedback. Second, we suppose the experimenter also processes each measurement record to have more detailed information about individual quantum state trajectories and so can choose whether a particular state is suitably prepared on demand. We include environmental effects in our analysis with parameters chosen to be consistent with current experimental technology for dispersive transmon trajectory measurements \cite{Kater2013,Weber2014,Roch2014,Hacohen-Gourgy2016}.

In the Markovian case with no signal filtering or feedback delay, we solve the two variations of the problem analytically and verify the results numerically. We find that in the ideal case with no environmental effects, both ways of posing the problem coincide: Any angle in the qubit Bloch sphere may be perfectly targeted by a suitable choice of linear feedback parameters. When environmental effects are added (neglecting feedback delay and signal filtering), we find that the two variations still coincide provided the feedback parameters are optimally chosen. That is, targeting any angle in the Bloch sphere with maximum ensemble-averaged purity is equivalent to stabilizing individual trajectories such that their measurement backaction is minimized. We find that all angles not too near the natural measurement pole corresponding to the excited qubit state may still be targeted with high purity, and solve for the feedback parameters needed to obtain each such state. Notably, while the ensemble-averaged purity remains nearly constant, the distributions of individual trajectories spread asymmetrically near the measurement poles, developing peaks of maximum probability with higher purity than the ensemble average. For stabilized states at the equator of the Bloch sphere, the maximum probability peak coincides with the ensemble-averaged state. For stabilized states near the excited state measurement pole, the maximum probability peak bifurcates into two distinct regions, with one dominant, such that each region is substantially more pure than the ensemble-average state. As a result, the two ways of posing the stabilization problem have distinct solutions for states near the natural measurement poles. For stabilized states near the ground state measurement pole, both the ensemble average and peak purify due to energy-relaxation environmental effects. 

Once signal filtering and feedback delay are included, the dynamics become intrinsically non-Markovian due to the delay buffer. We investigate these effects numerically. Compared to the Markovian case, we find that the stabilization is minimally degraded by single-pole (RC) low-pass filtering of the signal, so the protocol appears robust to realistic frequency-filtering effects in the feedback circuitry. We also find that feedback delay has the largest negative effect on the protocol, causing a dramatic reduction in stabilization purity as the delay becomes comparable to the measurement collapse timescale. This sensitivity to delay agrees with the feedback analysis performed for stabilized Rabi oscillations of a double-quantum-dot setup in Ref. \cite{Ruskov2004}. In our case, the trajectory distributions rapidly broaden with delay since the delayed signal being fed back into the controller has diminished relevance for the later state evolution that is being controlled. Nevertheless, the effect of delay can be mitigated by slowing the measurement rate to keep the collapse timescale an order of magnitude longer than the delay in the feedback circuitry. Curiously, despite the fact that in the Markovian case the Bloch sphere equator can be stabilized, we find that any amount of filtering or delay destabilizes a sharp angular region around the equator, which then becomes consistent with the instability observed at the equator in fluorescence-based protocols \cite{wangwiseman2001,Ibarcq2016}. In summary, most angles in the qubit Bloch sphere may be stabilized with high purity even when realistic models of the environment and feedback circuitry are taken into account.

The paper is organized as follows. In Sec.~\ref{sec:system}, we detail the modeling of the dispersive qubit monitoring with feedback for both idealized and realistic situations. In Sec.~\ref{sec:averagestate}, we pose the problem of ensemble-averaged qubit state stabilization and discuss both analytical results and numerical simulations. In Sec.~\ref{sec:trajectory}, we pose the related problem of state stabilization for individual qubit trajectories and discuss both analytical results and numerical simulations. We conclude in Sec.~\ref{sec:conclusion}.
	
\section{Feedback Model}\label{sec:system}

For concreteness, we consider a superconducting transmon qubit that is dispersively coupled to a microwave resonator as the measuring apparatus \cite{koch07,Gambetta2008,Korotkov2014,Korotkov2016}. The qubit Hamiltonian describing the lowest two levels of the transmon is $\hat{H}_q = \hbar \omega_q\,\hat{\sigma}_z/2$, where the Pauli operator $\hat{\sigma}_z = \ket{1}\!\bra{1}-\ket{0}\!\bra{0}$ indicates the difference between the excited ($\ket{1}$) and ground ($\ket{0}$) state energy levels, and $\omega_q$ is the natural qubit frequency. We will work in the rotating frame of the qubit in what follows for simplicity to neglect this natural evolution. The Hamiltonian of the coupled microwave resonator mode is harmonic $\hat{H}_r = \hbar\omega_r\,\hat{a}^\dagger\hat{a}$, where $\hat{a}$ is its lowering operator and $\omega_r$ is the natural resonator frequency that is detuned from the qubit. The dispersive coupling Hamiltonian between the qubit and resonator then approximates $\hat{H}_{qr} = \hbar\chi\,\hat{\sigma}_z\,\hat{a}^\dagger\hat{a}$, where $\pm\chi$ is the qubit-state-dependent frequency shift of the resonator due to the dispersive coupling. The resonator has energy-decay rate $\kappa$ and is pumped on resonance with a coherent field to reach a qubit-state-dependent coherent steady state with qubit-state-\emph{independent} average photon number $\bar{n}$. The steady-state microwave field then leaks from the resonator and travels down a transmission line, where it is amplified along the maximally informative quadrature \cite{Korotkov2014,Korotkov2016} with a phase-sensitive amplifier and finally measured to produce a time-continuous homodyne signal $I(t)$. Due to the entanglement with the leaked microwave field, definite qubit states $\ket{0}$ and $\ket{1}$ produce noisy homodyne signals that temporally average to distinct mean values $I_0$ and $I_1$, respectively. As such, monitoring the homodyne signal provides time-continuous information about the qubit energy basis. On average, these homodyne measurements decohere the qubit state at the ensemble-averaged measurement-dephasing rate $\Gamma_m = 8\chi^2\bar{n}/\kappa$ \cite{Gambetta2008,Korotkov2014,Korotkov2016}, which may be determined experimentally by comparing the Ramsey decay rates of the qubit with and without measurement \cite{Kater2013,Weber2014,Roch2014,chantasri2016}.

To stabilize the qubit state, we process the homodyne signal $I(t)$ and feed it back into the coherent qubit evolution as a coherent Rabi drive \cite{Wiseman1993feedback,Wiseman1994,Hofmann1998,Doherty1999,wangwiseman2001,Wiseman2002bayes,wang2005entfeed,Ibarcq2016}. For convenience, we first rescale the signal so that it directly corresponds to $\hat{\sigma}_z$. The rescaled readout has the form $r(t) \equiv 2(I(t) - \bar{I})/\Delta I$, where $\bar{I} = (I_0 + I_1)/2$ is the mean homodyne current, and $\Delta I = I_1 - I_0$ is the homodyne signal contrast. The rescaled readout $r(t)$ then has a temporal average of $\pm 1$ for definite qubit energy states, in agreement with the eigenvalues of $\hat{\sigma}_z$. Before being fed back into the controller, this rescaled readout is necessarily frequency-filtered by the feedback circuitry \cite{Ruskov2004}, $r \mapsto \tilde{r}$, which also delays the filtered readout by a duration $T_d$. This filtered and delayed readout $\tilde{r}(t-T_d)$ is then used to modulate a coherent Rabi drive described by the control Hamiltonian,
\begin{align}\label{eq:feedbackham}
  \hat{H}_c = \hbar\left[\Delta_0 + \Delta_1\,\tilde{r}(t-T_d)\right]\,\frac{\hat{\sigma}_\phi}{2},
\end{align}
where $\hat{\sigma}_\phi = \cos\phi\,\hat{\sigma}_x + \sin\phi\,\hat{\sigma}_y$, with $\hat{\sigma}_x = \ket{1}\!\bra{0}+\ket{0}\!\bra{1}$ and $\hat{\sigma}_y = -i\ket{1}\!\bra{0} + i\ket{0}\!\bra{1}$. The choice of angle $\phi$ fixes the plane of the qubit Bloch sphere to which the qubit will become stabilized. For simplicity of discussion, we now fix $\phi=\pi$ to choose clockwise oscillations in the $yz$-plane. This feedback Hamiltonian induces oscillations at a constant Rabi frequency $\Delta_0$ that is modulated by the linear feedback term $\Delta_1\,\tilde{r}(t-T_d)$. As we will show, and in agreement with fluorescence-based feedback stabilization protocols \cite{wangwiseman2001,Wiseman2002bayes,wang2005entfeed,Ibarcq2016}, the two parameters $\Delta_0$ and $\Delta_1$ completely control the effect of the feedback by jointly determining the attraction region in state space for stabilization. In contrast to the fluorescence-based protocols that have a bias toward the ground state, we will show that dispersive measurements can stabilize nearly any angle within the chosen plane with high purity (i.e., nearly any pure state may be targeted with high fidelity).

We now describe the qubit evolution produced by collecting the dispersive readout $r(t)$ and applying the feedback Hamiltonian in Eq.~\eqref{eq:feedbackham}. We first consider a time-discrete quantum Bayesian model \cite{Korotkov2014,Korotkov2016}, from which we can numerically simulate the feedback protocol including inefficiencies and delay. We then consider a time-continuous stochastic master equation model \cite{Gambetta2008} that interpolates the Markovian feedback dynamics in the limit of zero delay $T_d \to 0$ (and no signal filtering) with a fictitious time-continuous stochastic process \cite{Wiseman1993feedback,Wiseman1994,Hofmann1998,wangwiseman2001,Wiseman2002bayes,wang2005entfeed}, and which is convenient for analytic derivations of the required feedback parameters.

\subsection{Time-discrete model}
The strength of the qubit measurement is determined by the timescale $\tau_m$ needed to obtain a unit signal-to-noise ratio for distinguishing the qubit states. A qubit initially in a superposition of energy states will typically collapse to a definite energy state within a few $\tau_m$. This collapse timescale $\tau_m$ is related to the ensemble dephasing rate $\Gamma_m$ of the qubit due to measurement according to $\Gamma_m = (2\tau_m\eta)^{-1}$, where $\eta \in [0,1]$ is the quantum efficiency of the microwave detection circuitry \cite{Korotkov2014,Korotkov2016}. 

When the energy decay rate of the resonator $\kappa$ is sufficiently fast compared to both the measurement rate $\Gamma_m$ and the dynamics of the qubit (known as the ``bad cavity regime''), the coherently pumped resonator effectively remains at steady-state. We may then approximately ignore the resonator and phenomenologically describe the qubit evolution alone \cite{Korotkov2014}, provided that we coarse-grain the collected readout $r(t)$ into discrete segments $\{\bar{r}_k\}$ that are temporal averages $\bar{r}_k\,dt \equiv \int_{t_k}^{t_k+dt}r(t')dt'$ over time increments $dt > \kappa^{-1}$ longer than the relaxation timescale of the resonator field \cite{Korotkov2016}. This coarse-grained measurement evolution is then Markovian in the discretized time intervals, and can be efficiently described using the quantum Bayesian method \cite{Korotkov2014,Korotkov2016}, which also permits efficient numerical simulation.
	
The total evolution of the qubit can be approximately decomposed into three parts: pure measurement backaction, coherent unitary evolution, and environmental dissipation. These three contributions may be treated separately provided that the duration $dt$ is smaller than the characteristic timescales of each evolution. We now describe each contributing evolution, then combine them.

First, the form of the measurement backaction follows from Bayes' theorem and the Born rule \cite{Korotkov2014}. Given empirically measured conditional probabilities $P(\bar{r}|0)$ and $P(\bar{r}|1)$ for obtaining a result $\bar{r}$ averaged over $dt$ given a definite qubit energy state (0 or 1), the qubit energy probabilities must update according to Bayes' theorem $P(0)\mapsto P(0|\bar{r}) = P(\bar{r}|0)P(0)/P(\bar{r})$ and $P(1)\mapsto P(1|\bar{r}) = P(\bar{r}|1)P(1)/P(\bar{r})$, where the total probability for collecting $\bar{r}$ is $P(\bar{r}) = P(\bar{r}|0)P(0)+P(\bar{r}|1)P(1)$. This evolution is equivalent to the density operator update rule $\hat{\rho} \mapsto \hat{M}_{\bar{r}}\hat{\rho}\hat{M}^\dagger_{\bar{r}}/P(\bar{r})$ with probability $P(\bar{r}) = \text{Tr}(\hat{M}^\dagger_{\bar{r}}\hat{M}_{\bar{r}}\hat{\rho})$, determined entirely by the Kraus operator
\begin{align}
\hat M_{\bar{r}} = \sqrt{P(\bar{r}|0)}\, \ket{0}\!\bra{0} + \sqrt{P(\bar{r}|1)} \ket{1}\!\bra{1}.
\end{align}
For a maximally informative choice of amplified quadrature, there is no additional phase-backaction on the qubit \cite{Korotkov2014,Korotkov2016}, so this minimal Kraus operator is sufficient.

Because of the central limit theorem, the conditional probabilities $P(\bar{r}|0)$ and $P(\bar{r}|1)$ approximate Gaussians centered at $\pm 1$. For simplicity, we also assume approximately equal variances, which may be justified empirically. To preserve proper scaling of temporal averages as the discrete time step $dt$ is varied, the variances of each time step must be inversely proportional to $dt$. By definition of the unit signal-to-noise timescale $\tau_m$, the variances must then have the form $\tau_m/dt$, yielding 
\begin{align}\label{eq:measop}
\hat M_{\bar{r}} &= C(\bar{r})\left[ e^{-\bar{r} dt/2\tau_m} \ket{0}\!\bra{0} + e^{\bar{r} dt/2\tau_m} \ket{1}\!\bra{1} \right] 
\end{align}
where $C(\bar{r}) = (dt/2 \pi \tau_m)^{1/4}e^{-(\bar{r}^2+1) dt/4 \tau_m}$ is a state-independent normalization factor that cancels in the state update rule. Note that this Kraus operator may be conveniently written $\hat{M}_{\bar{r}} = C(\bar{r})\exp(\bar{r}dt\hat{\sigma}_z/2\tau_m) = C(\bar{r})[\cosh(\bar{r}dt/2\tau_m)\hat{1} + \sinh(\bar{r}dt/2\tau_m)\hat{\sigma}_z]$, with $\hat{1} = \ket{0}\!\bra{0} + \ket{1}\!\bra{1}$. The rate $(2\tau_m)^{-1}$ that appears here is the ensemble-average dephasing rate obtained after averaging over all collected signal $\bar{r}$.

Second, the form of the unitary dynamics follows from the control Hamiltonian $\hat{H}_c$ in Eq.~\eqref{eq:feedbackham} with $\phi=\pi$, which describes precession in the $yz$-plane of the qubit Bloch sphere. Treating the coherent evolution independently yields the unitary operator $\hat{U} = \exp(-idt\hat{H}_c/\hbar)$ that simplifies to the convenient form
\begin{align}\label{eq:unop}
\hat{U} &= e^{idt\Delta\,\hat{\sigma}_x/2} = \cos(\Delta\,dt/2)\hat{1} + i\sin(\Delta\,dt/2)\hat{\sigma}_x
\end{align}
with a frequency $\Delta = \Delta_0 + \Delta_1\,\tilde{\bar{r}}(t-T_d)$ that depends upon the filtered average readout $\tilde{\bar{r}}(t-T_d)$ collected from a time delay $T_d=n_d\,dt$ of $n_d$ time steps in the past and fed back into the controller during the current time interval $dt$. The nonzero feedback delay makes the resulting dynamics intrinsically non-Markovian. Additional frequency-filtering from the feedback circuitry \cite{Ruskov2004} is discretely modeled by a recursive transformation of the history of raw average readouts $\{\bar{r}_k\}$ stored in a memory buffer (see Sec.~\ref{sec:delay} for more detail). 

Third, we consider the most common forms of environmental dissipation: energy relaxation, energy dephasing, and detector inefficiency. Energy relaxation follows a phenomenological decay rate $1/T_1$ for the excited state to relax into the ground state. Defining the density operator matrix elements as $P_0 = \bra{0}\hat{\rho}\ket{0}$, $P_1 = \bra{1}\hat{\rho}\ket{1}$, and $\rho_{01} = \bra{0}\hat{\rho}\ket{1} = \rho_{10}^* \propto \sqrt{P_0 P_1}$, energy decay yields the transformation $P_1 \mapsto P_1 \exp(-dt/T_1)$, which forces $P_0 \mapsto 1-P_1\exp(-dt/T_1)$ and $\rho_{01}\mapsto\rho_{01}\exp(-dt/2T_1)$. We summarize these rules with the energy-decay map $\hat{\rho}\mapsto\mathcal{E}_{T_1}(\hat{\rho})$. Similarly, environmental energy dephasing follows a phenomenological decay rate $1/T_2$, and corresponds to the transformation $\rho_{01} \mapsto \rho_{01}\exp(-dt/T_2)$, summarized by the dephasing map $\hat{\rho}\mapsto\mathcal{E}_{T_2}(\hat{\rho})$. The detector inefficiency $(1-\eta)$ describes the fraction of measured qubit information that has been lost to the environment, thus producing residual measurement dephasing on average with rate $\gamma = \Gamma_m - 1/2\tau_m = (1-\eta)/(2\tau_m\eta)$, which is summarized by the map $\hat{\rho}\mapsto\mathcal{E}_\gamma(\hat{\rho})$ that dephases in a similar way to $\mathcal{E}_{T_2}$; i.e., $\rho_{01} \mapsto \rho_{01}\exp(-\gamma dt)$. 

The total state update is then approximately described by the composite map
\begin{align}\label{eq:update}
\hat{\rho}(t+dt) = (\mathcal{E}_\gamma\circ\mathcal{E}_{T_2}\circ\mathcal{E}_{T_1})\left[\frac{ {\hat U} {\hat M}_{\bar{r}} \hat{\rho}(t) {\hat M}_{\bar{r}}^{\dagger} {\hat U}^{\dagger}}{{\rm Tr}\left( { {\hat M}_{\bar{r}}^{\dagger}\hat M}_{\bar{r}} \hat{\rho}(t) \right)}\right],
\end{align}
where all operations are described at time $t$. This separation of the evolution into distinct pieces is formally valid to linear order in $dt$, but (unlike explicitly linear-order updates) ensures completely positive evolution of the state. 

To minimize accumulated error from the composition approximation in Eqs.~\eqref{eq:pureupdate} and \eqref{eq:update}, the time step should satisfy $dt/\tau_m \ll 1$, $dt\,\Delta \ll 1$, and $dt\,\Gamma_m \ll 1$. In particular, since the noisy feedback signal $\tilde{\bar{r}}(t-T_d)$ will likely dominate the rate $\Delta = \Delta_0 + \Delta_1 \tilde{\bar{r}}(t-T_d)$, it should satisfy $dt\,\tilde{\bar{r}}\,\Delta_1 \ll 1$. Since $\bar{r} \lesssim 5\,\sqrt{\tau_m/dt}$ by Gaussian statistics, this implies $dt\,\tilde{\bar{r}}\,\Delta_1 < 5\Delta_1\sqrt{dt\,\tau_m} \ll 1$, so $\Delta_1 \ll 1/(5\sqrt{dt\,\tau_m})$ is a practical upper bound for $\Delta_1$ for a chosen time step $dt$ in Eq.~\eqref{eq:update}. We also note that the chosen ordering of operations, in particular measurement backaction followed by feedback control, is made to anticipate the Markovian feedback limit with vanishing delay ($T_d \to 0$) in the next section.

To efficiently simulate Eq.~\eqref{eq:update} numerically, we use a Bloch coordinate decomposition of the state, $\hat{\rho} \to (x,y,z)$, where $x = \text{Tr}\left(\hat{\sigma}_x\,\hat{\rho}\right)$, $y = \text{Tr}\left(\hat{\sigma}_y\,\hat{\rho}\right)$, and $z = \text{Tr}\left(\hat{\sigma}_z\,\hat{\rho}\right)$. With this representation, the rescaled average readout $\bar{r}$ at each time $t$ may be sampled from the readout distribution
\begin{align}\label{eq:readout}
  P(\bar{r}) = \text{Tr}\left(\hat{M}_{\bar{r}}^\dagger\hat{M}_{\bar{r}}\hat{\rho}\right) &= P_0\,P(\bar{r}|0)+P_1\,P(\bar{r}|1) \nonumber \\ &\approx \frac{\exp(-dt(\bar{r}-z)^2/2\tau_m)}{\sqrt{2\pi\tau_m/dt}}.
\end{align}
where the single-Gaussian approximation makes numerical sampling efficient. After generating each $\bar{r}$, the update in Eq.~\eqref{eq:update} is applied, using the previously sampled, time-delayed, and filtered $\tilde{\bar{r}}$ as appropriate in the feedback unitary. 

Expressed in Bloch-coordinates, Eq.~\eqref{eq:update} has a simple prescription: Given the state $(x_n,y_n,z_n)$ and a random $\bar{r}$ sampled at time step $t_n$, as well as a filtered $\tilde{\bar{r}}_{n-n_d}$ from time step $t_n-T_d$ with $T_d = n_d\, dt$, the new state $(x_{n+1},y_{n+1},z_{n+1})$ at time $t_{n+1}=t_n+dt$ is obtained by the following sequence of transformations
\begin{align}\label{eq:blochupdate}
  p_n &= \cosh(\bar{r}dt/\tau_m) + z_n\,\sinh(\bar{r}dt/\tau_m), \nonumber \\
  x_n' &= x_n/p_n, \nonumber \\ 
  y_n' &= y_n/p_n, \nonumber \\
  z_n' &= [z_n\,\cosh(\bar{r}dt/\tau_m) + \sinh(\bar{r}dt/\tau_m)]/p_n, \nonumber \\
  \Delta &= \Delta_0 + \Delta_1\,\tilde{\bar{r}}_{n-n_d}, \nonumber \\
  y_n'' &= y_n'\,\cos(dt\Delta) + z_n'\,\sin(dt\Delta), \nonumber \\
  z_n'' &= z_n'\,\cos(dt\Delta) - y_n'\,\sin(dt\Delta), \nonumber \\
  x_{n+1} &= x_n'\, e^{-dt/2T_1 -dt/T_2 - dt(1-\eta)/2\tau_m\eta}, \nonumber \\
  y_{n+1} &= y_n''\, e^{-dt/2T_1 -dt/T_2 - dt(1-\eta)/2\tau_m\eta}, \nonumber \\
  z_{n+1} &= z_n''\,e^{-dt/T_1} - (1-e^{-dt/T_1}),
\end{align}
Note that the normalization factor $p_n$ is proportional to the probability for sampling $\bar{r}$, but irrelevant constants have been canceled. At the start of each simulated evolution, there is no pre-sampled readout history, so we treat the applied feedback $\tilde{\bar{r}}_{n-n_d}$ as zero in the update until the feedback delay buffer has been filled. For simplicity, we set $x=0$ initially, which constrains the dynamics to the $yz$-plane.

\subsection{Time-continuous model}
For analytic convenience, the time-continuum limit $dt \to 0$ can be used to derive stochastic master equations (SMEs) from the update equation Eq.~\eqref{eq:update} in the Markovian feedback limit with $T_d \to 0$. Note that this limit formally produces a physically incorrect model for two reasons: First, $dt > \kappa^{-1}$ must be satisfied to phenomenologically ignore the relaxation of the resonator to its steady state. The continuum limit thus implies a resonator with infinitely fast energy decay, $\kappa\to\infty$, which would produce no qubit measurement, since $\Gamma_m \propto \kappa^{-1} \to 0$. Second, $T_d > 0$ in any realistic laboratory feedback control loop due to finite signal velocity and controller processing time, so the limit $T_d \to 0$ is unrealizable. Nevertheless, this Markovian feedback continuum limit still formally produces a useful interpolation of the discretized dynamics in the preceding section as an idealized time-continuous stochastic process that may be more easily analyzed \cite{Gambetta2008}. That is, averaging this fictitious process over time bins $dt > \kappa^{-1}$ should still recover the physically correct time-discretized evolution. In particular, it is simple to obtain a master equation (ME) describing the ensemble-averaged behavior from such an SME. It is also a good approximation when the delay is short compared to the intrinsic collapse time of the measurement, $T_d \ll \tau_m$. Indeed, such a separation of time scales is always implied by such time-continuous stochastic equations of motion \cite{BookGardiner2,BookCarmichael}.

Stochastic processes are not differentiable. Thus, a careful treatment is needed to derive the differential equations for the evolution \cite{BookGardiner2}. There are two popular pictures for stochastic differential equations, It\^o and Stratonovich, the former having the advantage that the stochastic noise at any time $t$ is independent from the process in which it occurs (see Ref. \cite{BookGardiner2} and Appendix B of Ref. \cite{BookWiseman}). This advantage makes finding the ME for ensemble averaged evolution straightforward in the It\^o picture, so we opt for that approach here.

Following the It\^o formalism, we observe that Eq.~\eqref{eq:readout} may be written as $P(\bar{r}) =  P(dW)dt/\sqrt{\tau_m}$, with the zero-mean Gaussian distribution $P(dW) = \exp(-dW^2/2dt)/\sqrt{2\pi dt}$ having variance $dt$, provided that we make the identification $\bar{r}\,dt = z\,dt + \sqrt{\tau_m}\,dW$. The random variable $dW$ then has the properties of a Wiener increment, so that the identity $dW^2 = dt$ becomes exact in the continuum limit $dt\to 0$ \cite{BookGardiner2}. As such, we can interpolate the coarse-grained time-dependent readout as a moving-mean stochastic process
\begin{align}
  \bar{r}(t) = z(t) + \sqrt{\tau_m}\, \xi(t)
\end{align}
centered at $z(t)$ with additive white noise $\xi(t) = dW/dt$ with zero-mean $\langle \xi(t) \rangle =0$ and Dirac-$\delta$ temporal-correlation $\langle \xi(t) \xi(t') \rangle = \delta(t-t')$. 

We can now expand the right hand side of the update equation Eq.~\eqref{eq:update} to first order of $dt$ while adhering to the It\^o rule $dW^2 = dt$, and assuming $T_d \to 0$ (implying the filtered readout $\tilde{\bar{r}}$ in $\hat{U}$ becomes identified with the just-sampled and unfiltered $\bar{r}$ in $\hat{M}_{\bar{r}}$). This expansion produces the Markovian feedback It\^o stochastic master equation
\begin{align}\label{eq:feedbacksme}
d\hat{\rho} &= i\frac{\Delta_0}{2}[\hat{\sigma}_x,\,\hat{\rho}]\,dt +i\frac{\Delta_1}{4}[\hat{\sigma}_x,\,\{\hat{\sigma}_z,\,\hat{\rho}\}]\,dt \nonumber \\
&+ \frac{\Gamma'}{2}\mathcal{D}[\hat{\sigma}_z]\hat{\rho}\,dt + \frac{1}{T_1}\mathcal{D}[\hat{\sigma}_-]\hat{\rho}\,dt 
+ \frac{\tau_m\,\Delta_1^2}{4}\mathcal{D}[\hat{\sigma}_x]\hat{\rho}\,dt \nonumber \\
& + \frac{1}{2}\mathcal{H}[\hat{\sigma}_z]\hat{\rho}\,\frac{dW}{\sqrt{\tau_m}} +i\frac{\tau_m\,\Delta_1}{2}[\hat{\sigma}_x,\,\hat{\rho}]\,\frac{dW}{\sqrt{\tau_m}}.
\end{align}
Here $\Gamma' = 1/T_2 + 1/2\tau_m\eta$ and $\hat{\sigma}_- = \ket{0}\!\bra{1}$. We also define the dissipation superoperator $\mathcal{D}[\hat{A}]\hat{\rho} = \hat{A}\hat{\rho}\hat{A}^\dagger - \{\hat{A}^\dagger\hat{A},\,\hat{\rho}\}/2$, as well as the innovation superoperator $\mathcal{H}[\hat{A}]\hat{\rho} = \hat{A}\hat{\rho} + \hat{\rho}\hat{A}^\dagger - \text{Tr}(\hat{A}\hat{\rho} + \hat{\rho}\hat{A}^\dagger)\hat{\rho}$ \cite{BookWiseman}. 
Of particular importance is the commutator of the anti-commutator that appears in the first line, which describes the effect of the feedback control applied immediately after the collapse induced by the measurement \cite{Wiseman1993feedback,Wiseman1994,BookWiseman}. The ensemble-averaged ME is obtained simply by setting the terms proportional to $dW$ to zero in Eq.~\eqref{eq:feedbacksme}.

This result simplifies when written in terms of the Bloch coordinates, yielding
	\begin{subequations}\label{eq:nonidealME}
		\begin{align}
		\dot{x} &= -\Gamma\, x - x\, z \frac{\xi}{\sqrt{\tau_m}} ,\\
		\dot{y} &= -\left[\Gamma + \frac{\tau_m\,\Delta_1^2}{2}\right] y + \Delta_0 z + \Delta_1 \nonumber \\
        &\qquad - y\,z\,\frac{\xi}{\sqrt{\tau_m}} + \tau_m\, \Delta_1\, z\, \frac{\xi}{\sqrt{\tau_m}} ,\\
		\dot{z} &= -\frac{\tau_m\,\Delta_1^2}{2}\,z- \Delta_0 y  - \frac{1+z}{T_1} \nonumber \\
        &\qquad + (1-z^2)\,\frac{\xi}{\sqrt{\tau_m}} - \tau_m\,\Delta_1\, y\, \frac{\xi}{\sqrt{\tau_m}},
		\end{align}
	\end{subequations}
where $\Gamma = 1/2 T_1 + 1/T_2 + 1/2 \tau_m\eta $ is the total ensemble $z$-dephasing rate in the absence of feedback. Note that the feedback contributes additional $x$-dephasing at a rate $\tau_m\Delta_1^2/2$. The ensemble-averaged ME can be recovered from this SME by setting the terms proportional to $\xi$ to zero. The consistency between the deterministic ensemble-average ME and the discrete formulation of Eq.~\eqref{eq:update} is then checked by averaging several thousand trajectories numerically generated using Eq.~\eqref{eq:update} and comparing to the analytic solution of the deterministic part of Eq.~\eqref{eq:nonidealME}. We will primarily use the Bloch coordinate representation in what follows for analytical calculations.

\section{Average State Stabilization}\label{sec:averagestate}

We now pose the following state stabilization problem:
Consider a closed, continuous measurement feedback loop that is to be used as a storage mechanism for preparing a target qubit state on demand. An experimenter will press a button at a unknown future time, terminating the feedback loop, in order to receive the prepared qubit state for immediate experimentation. \emph{Can the feedback parameters $\Delta_0$ and $\Delta_1$ be fixed such that a particular qubit state is reliably prepared on average?} 

We show that in the case of idealized Markovian feedback an arbitrary pure state may be prepared through feedback alone. We then generalize the protocol by adding realistic experimental nonidealities, including energy decay, environmental dephasing, and measurement inefficiency, showing that the best prepared states have degraded purity compared to the ideal case, but may still be meaningfully stabilized by the feedback loop. Finally, we discuss the significant effect of realistic signal filtering and feedback delay on the stabilization protocol, and show that good stabilization is still possible when the delay is sufficiently short compared to the measurement collapse timescale. 

\subsection{Stationary states for ideal Markovian feedback}\label{sec:ideal}

\begin{figure}[t]
	\includegraphics[width=0.9\columnwidth]{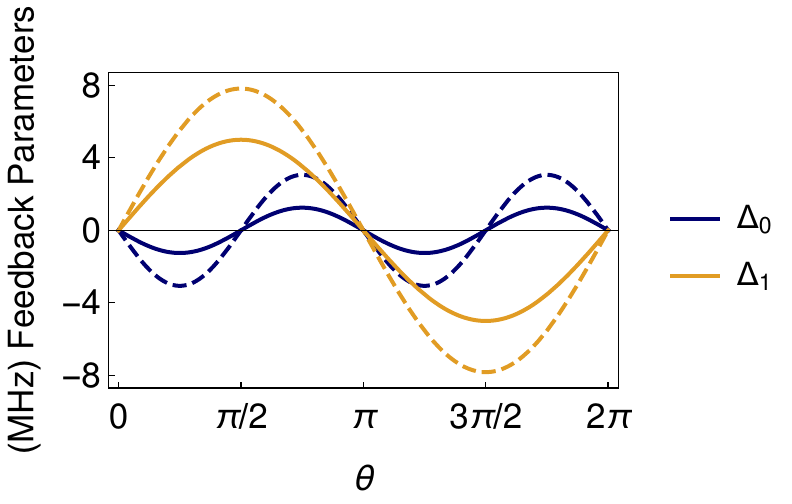}
	\caption{ Markovian feedback parameters for ensemble-average stabilization. The constant Rabi drive strength $\Delta_0$ and linear feedback parameter $\Delta_1$ are shown as functions of the stabilized Bloch sphere polar angle $\theta$ in the $yz$-plane, in units of $\tau_m^{-1}$ with $\tau_m = 0.2 \mu s $. (Solid) Ideal stabilization of a pure state. (Dashed) Nonideal stabilization of the state with maximum Bloch radius $R_{\rm max} = 1/\sqrt{1/\eta + 2\tau_m/T_2} = 0.64$ which includes effects of environmental dephasing ($T_2 = 40$ $\mu$s) and reduced quantum efficiency ($\eta = 0.41$). }
	\label{fig:feedbackpm}
\end{figure}

For the case of a pure state ($T_1,T_2\to\infty, \, \eta = 1$) with no feedback delay ($T_d \to 0$), we use the Bloch Eqs.~\eqref{eq:nonidealME} to answer the question posed above. Specifically, consider the case where $x=0$ initially, so the qubit state lies in the $yz$-plane at $(0,y_s,z_s)$, with $y_s = \sin\theta_s$ and $z_s = \cos\theta_s$ determined entirely by the polar angle in the Bloch sphere. We then demand the condition that this initial state be a fixed point of the ensemble-averaged dynamics, meaning that $\dot{y}=\dot{z}=0$ in Eq.~\eqref{eq:nonidealME} with $\xi \to 0$.
Solving this constraint yields the feedback parameters
\begin{align}\label{eq:idealparams}
  \Delta_0 &= -\frac{y_sz_s}{2\tau_m} = -\frac{\sin2\theta_s}{4\tau_m}, & \Delta_1 &= \frac{y_s}{\tau_m} = \frac{\sin\theta_s}{\tau_m}.
\end{align}
That is, for any choice of state angle $\theta_s$, there exist unique feedback parameters $\Delta_0$ and $\Delta_1$ that stabilize that state on average. These parameters are shown as functions of the stabilized state angle $\theta_s$ in Figure~\ref{fig:feedbackpm}. The dependence of the linear feedback parameter $\Delta_1$ on $\sin\theta_s$ is sensible, since for $\theta_s = \{0,\pi\}$ the natural measurement poles are already fixed points for collapse. Similarly, it is sensible for the constant rate $\Delta_0$ to have dependence on $\sin 2\theta_s$, since the constant rotation biases the fixed point asymmetrically and should vanish for the backaction-symmetric stabilization points on the equator with $\theta_s = \pm\pi/2$.

Numerical simulations are in excellent agreement with this result for the ideal stationary state, with the feedback consistently isolating a unique stabilized state. The only exception to the uniqueness of the stabilization is that the active control vanishes at the natural measurement poles, since $\Delta_0,\, \Delta_1 \rightarrow 0$ as $\theta_s \rightarrow 0,\, \pi$. The absence of feedback at the poles permits stochastic collapse to either $\hat{\sigma}_z$-eigenstate. This limitation can be readily ameliorated in practice, however, by stabilizing a state near the pole of choice before removing feedback and permitting a final collapse step with high probability to the desired pole, which yields a protocol analogous to standard state heralding.

\subsection{Stationary states for Markovian feedback}\label{sec:nonideal}

For realistic laboratory situations, we must add the effects of the energy decay time $T_1$, dephasing time $T_2$, and imperfect measurement efficiency $\eta$. These non-idealities decrease the state purity, so states in the $yz$-plane will be characterized by both a polar angle $\theta$ and a radius $R$, related to the purity by $\text{Tr}(\hat{\rho}^2) = (1+R^2)/2$, such that $y = R\sin\theta$ and $z=R\cos\theta$. Since a pure state is characterized entirely by the polar angle $\theta$, we may also consider the radius $R$ to be a measure of the average \emph{state fidelity} 
\begin{align}
   F=\bra{\theta_s}\hat{\rho}\ket{\theta_s} = \frac{1+R}{2}
\end{align}
between a target pure state and a mixed state at the same stabilized angle $\theta_s$. We will see that the stabilized angle $\theta_s$ may be reliably targeted, which makes $R$ a sensible measure of stabilization quality (as an aside, it is equivalent to the ``synchronization degree'' $D$ used in Ref.~\cite{Korotkov2005}).

As before, we demand that $\dot{y}=\dot{z}=0$ in Eq.~\eqref{eq:nonidealME} for the ensemble average when $\xi\to 0$. This stability condition produces the following set of stationary states that correspond to each particular choice of feedback parameters $\Delta_0$ and $\Delta_1$
\begin{align}\label{eq:nonidealss}
y_s 
&= \frac{\Delta_1\left[\frac{\tau_m\Delta_1^2}{2}\right] + \frac{1}{T_1}\left[\Delta_1-\Delta_0\right]}{\Delta_0^2 + \left[\frac{1}{T_1} + \frac{\tau_m \Delta_1^2}{2}\right]\left[\Gamma + \frac{\tau_m \Delta_1^2}{2}\right]} \nonumber \\
z_s 
&= -\frac{\Delta_0\Delta_1 + \frac{1}{T_1}\left[\Gamma + \frac{\tau_m\Delta_1^2}{2} \right]}{\Delta_0^2 + \left[\frac{1}{T_1} + \frac{\tau_m \Delta_1^2}{2}\right]\left[\Gamma + \frac{\tau_m \Delta_1^2}{2}\right]}.
\end{align}

However, not every state in the $yz$-plane may be stabilized due to constraints on the average purity. This fact becomes more clear when we express the stationary states in polar form with $\tan\theta_s=y_s/z_s$ and $R_s^2 = y_s^2 + z_s^2$,
\begin{align}\label{eq:nonidealsspolar}
\tan\theta_s &= -\frac{\frac{\tau_m\Delta_1^2}{2\Delta_0} + \frac{1}{T_1}\frac{\Delta_1-\Delta_0}{\Delta_0\Delta_1}}{1 + \frac{1}{T_1\Delta_1}\left[\frac{\Gamma}{\Delta_0} + \frac{\tau_m\Delta_1^2}{2\Delta_0} \right]} \nonumber \\
&\xrightarrow{T_1\to\infty} - \frac{\tau_m\Delta_1^2}{2\Delta_0}, \nonumber \\
R_s &= \frac{|\sec\theta_s|\left|\Delta_0\Delta_1 + \frac{\Gamma}{T_1} + \frac{\tau_m\Delta_1^2}{2T_1} \right|}{\Delta_0^2 + \left[\frac{1}{T_1} + \frac{\tau_m \Delta_1^2}{2}\right]\left[\Gamma + \frac{\tau_m \Delta_1^2}{2}\right]} \nonumber\\
&\xrightarrow{T_1\to\infty} 
\frac{|\Delta_1|}{|\Delta_0||\sec\theta_s| + \Gamma|\sin\theta_s|} 
\end{align}
The negligible energy-decay limit $T_1\to\infty$ is simpler to analyze, yet still corresponds to a reasonable approximation (especially since its dephasing effects at rate $(2T_1)^{-1}$ can be formally included into the definition of $T_2$). In this limit, we can readily extremize the value of $R_s$ while keeping $\theta_s$ fixed in Eqs.~\eqref{eq:nonidealsspolar}, which is accomplished by fixing the ratio $\Delta_1^2/\Delta_0$ and then varying only $\Delta_1$. This procedure determines the maximum radius to be a constant, $R_{\rm max} = 1/\sqrt{1/\eta + 2\tau_m/T_2} < 1$, that is achieved when $\Delta_1 = \sin\theta_s/(\tau_m R_{\rm max})$. This upper bound on the radius indicates the maximum achievable purity for the set of stationary states, and thus the maximum achievable pure state preparation fidelity.

To set the control parameters to target a particular state $(y_s, z_s)$, we return to the $\dot{y}=\dot{z}=0$ condition for Eqs.~\eqref{eq:nonidealME} (remembering that we are dealing with the ensemble-average case where $\xi\to 0$) and solve for $\Delta_0$ and $\Delta_1$ directly to find
\begin{align}\label{eq:stationaryfeedback}
  \Delta_0 &= - \frac{\tau_m\Delta_1^2}{2}\frac{z_s}{y_s} - \frac{1+z_s}{T_1\,y_s}, \\
  \Delta_1 &= \frac{y_s}{R_s^2\tau_m}\Bigg[1 \pm  \sqrt{1-2\tau_mR_s^2\left[\Gamma+\frac{(1+z_s)z_s}{T_1\,y_s^2}\right]}\,\Bigg]. \nonumber
\end{align}
The $\Delta_1$ solution is not unique, but correctly becomes unique and reduces to Eqs.~\eqref{eq:idealparams} in the limits $T_1,T_2\to\infty$, and $\eta,R_s\to 1$. Moreover, in the $T_1\to\infty$ limit the condition for the maximum radius $R_{\rm max}$ derived above corresponds precisely to unique solutions of $\Delta_1$ (i.e., when the square root in $\Delta_1$ vanishes). It thus seems reasonable to hypothesize that the condition for maximum purity corresponds to unique solutions for $\Delta_1$ more generally, which constrains the radius to 
\begin{align}\label{eq:maxR}
	R_{\rm max}(\theta) &= \frac{1}{\frac{\tau_m}{T_1}\frac{\cos\theta}{\sin^2\theta} + \sqrt{\frac{2\tau_m}{T_1}(T_1\,\Gamma + \cot^2\theta) + \left[\frac{\tau_m}{T_1}\frac{\cos\theta}{\sin^2\theta}\right]^2}} \nonumber \\
	&\xrightarrow{T_1\to \infty} \frac{1}{\sqrt{2\tau_m\Gamma}} = \frac{1}{\sqrt{\frac{1}{\eta} + \frac{2\tau_m}{T_2}}}. 
\end{align}
We will revisit why this is the correct condition later when we analyze the noise-disturbance of individual trajectories in Sec.~\ref{sec:nonidealtraj}. Numerical simulations confirm that $R_{\rm max}(\theta)$ is the correct angle-dependent maximum radius that includes $T_1$. A plot of $R_{\rm max}(\theta)$ is shown in Fig.~\ref{fig:RpRe} of Sec.~\ref{sec:delaytraj}.

The angle-dependent $T_1$ corrections become significant only when $T_1 \sim \tau_m$, except around the poles. That is, for sufficiently strong measurements (shorter $\tau_m$), the effects of $T_1$ can be almost safely neglected. However, for an angular range between $\delta\theta = \pm\arcsin\sqrt{2\tau_m/T_1}$ near the poles, $T_1$ has a significant effect: It prevents stabilization of the excited state pole, while enhancing the stability of the ground state pole (see Fig~\ref{fig:RpRe}). Outside of this narrow angular range near each measurement pole, the limit of $\tau_m/T_1 \to 0$ should be adequate to describe most experiments.

For the $T_1\to \infty$ limit, the required control parameters have a simple closed form in terms of only the angle $\theta$ in the $yz$-plane
\begin{align}\label{eq:nonidealparams}
  \Delta_0 &= -\frac{\tau_m\Delta_1^2}{2\tan\theta} - \frac{1 + R_{\rm max}(\theta)\cos\theta}{T_1 R_{\rm max}(\theta)\sin\theta} \nonumber \\
  &\xrightarrow{T_1\to\infty} -\frac{\sin2\theta}{4\tau_m}\left[\frac{1}{\eta}+\frac{2\tau_m}{T_2}\right]  \nonumber \\ 
  \Delta_1 &= \frac{\sin\theta}{R_{\rm max}(\theta)\tau_m} 
  \xrightarrow{T_1\to\infty} \frac{\sin\theta}{\tau_m}\sqrt{\frac{1}{\eta} + \frac{2\tau_m}{T_2}},
\end{align}
and generalize Eq.~\eqref{eq:idealparams} in a straightforward way. This generalization is one of our main results, and is also shown in Fig.~\ref{fig:feedbackpm} for completeness.

\subsection{Non-Markovian experimental considerations}\label{sec:delay}

A fundamental assumption of our analytical treatment to this point has been the Markovian nature of the feedback implicit in the Bloch Eqs.~\eqref{eq:nonidealME}, which implies an infinitesimal feedback delay $T_d \to 0$ and thus no signal filtering. Although approximating the system time as continuous with no feedback delay yields the closed-form solutions in Eqs.~\eqref{eq:idealparams} and \eqref{eq:nonidealparams} for the stabilization parameters $\Delta_0$ and $\Delta_1$, the validity of these prescriptions may not hold for more realistic experimental situations (e.g., Ref. \cite{Ibarcq2016}). 

To address this shortcoming and more accurately model experiment, the effects of signal filtering and feedback delay within the feedback loop should be included. Signal filtering necessarily arises physically as a result of the finite bandwidth of the feedback circuitry. Feedback delay necessarily results both from the limited signal velocity in the feedback circuitry and any additional signal processing time within the feedback loop. Anticipating the need for these effects, we have already included them in the time-discrete model outlined in Eqs.~\eqref{eq:update} and \eqref{eq:blochupdate}. 

We model the finite circuitry bandwidth as a single-pole (RC) low-pass filter with exponential time constant $T_s$. That is, the output filtered signal at time step $k$ is an exponential moving average $\tilde{\bar{r}}_k = (dt/T_s) \sum_{j\leq k} \bar{r}_j \exp[-(k-j)dt/T_s]$ of the preceding collected raw signals $\{\bar{r}_j\}_{j\leq k}$. This moving average has the convenient recursive form
\begin{align}
  \tilde{\bar{r}}_k &= \tilde{\bar{r}}_{k-1} + (1 - e^{-dt/T_s})(\bar{r}_k - \tilde{\bar{r}}_{k-1}).
\end{align}
In the limit $T_s \to 0$ the bandwidth becomes infinite and the signal fed back into the control loop will be identical to the measured noisy signal. As the decay constant $T_s$ increases, the noisy signal will become increasingly smooth and lagged compared to the raw signal. As such, we expect the quality of the stabilization to degrade with increasing $T_s$, since the control signal will correspond less to the instantaneous qubit state. We consider the effect of bandwidths ranging from an ideal decay constant $T_s \to 0$ to the measurement collapse timescale $T_s \to \tau_m$.

We model the feedback delay $T_d = n_d\,dt$ by buffering $n_d$ time steps $dt$ of the filtered signal before feeding that buffer back into the unitary control in Eqs.~\eqref{eq:blochupdate}. As such, the effective Rabi frequency for the unitary control at time step $t_k$ is $\Delta_0 + \Delta_1\,\tilde{\bar{r}}_{k-n_d}$. Typical values of $T_d$ in current laboratory setups are $\sim\!200$--$500$ns \cite{Korotkov2012,Ibarcq2016}. As noted earlier, the lag in the feedback signal due to both signal filtering and delay makes the dynamics intrinsically non-Markovian, which largely precludes better analytical estimates of the optimal stabilization parameters. Nevertheless, we numerically explore the effects below.

\begin{figure}[t]
    \includegraphics[width=\columnwidth]{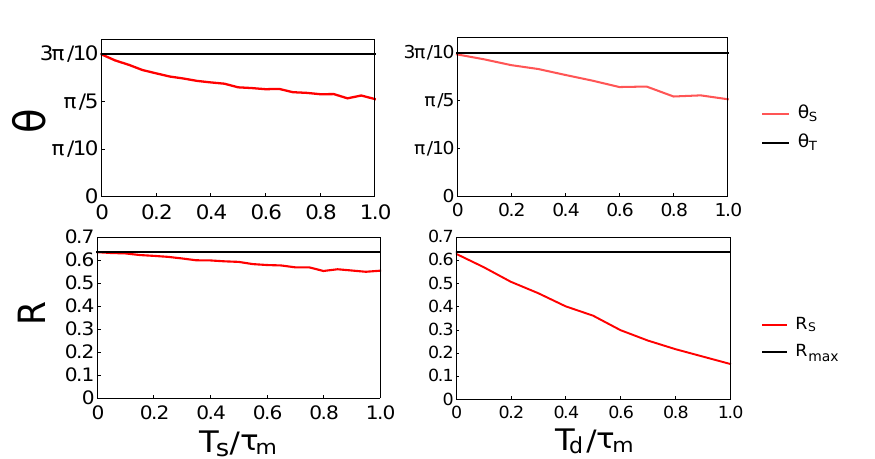}
	\caption{Effect of signal filtering and feedback delay on an ensemble-average stabilized state. Stabilized polar Bloch angle $\theta_s$ compared to the (representative) target angle $\theta_T=3\pi/10$ (upper) and Bloch radius $R_s$ compared to the target maximum radius $R_{\rm max}=0.64$ (lower) as functions of the exponential time constant $T_s/\tau_m$ for a low-pass filtered signal (left) and signal delay $T_d/\tau_m$ (right), both normalized by the measurement collapse time $\tau_m$. Though both effects cause a fixed and correctable angular drift relative to the target value, the feedback delay dramatically lowers the achievable purity for the stabilized state.}
    \label{fig:delay}
\end{figure}

As shown in the left column of Fig.~\ref{fig:delay}, the effect of signal filtering on the quality of the stabilized state are small. For a (representative) targeted state with angle $\theta_T = 3\pi/10$, the stabilized angle $\theta_s$ drifts nearly linearly by a small amount $\pi/10$ closer to the nearest measurement pole over the large filtering range $T_s\in[0,\tau_m]$. The stabilized Bloch radius $R_s$ is degraded from a target value $R_{\rm max} = 0.64$ by an amount $0.1$ over the same range of $T_s$. Notably, the radius remains essentially unaffected for $T_s \lesssim 0.2\tau_m$. Since the radius is closely related to the state purity, this robustness to small amounts of signal filtering indicates that quality state stabilization may still be achieved after a suitable recalibration of the targeted stabilization angle. Such correction of the angular drift will also compensate for most of the degradation in $R$ observed here. 

As shown in the right column of Fig.~\ref{fig:delay}, the effect of feedback delay on the stabilization quality is dramatic compared to the effect of signal filtering. While the (representative) target angle $\theta_T = 3\pi/10$ drifts by approximately the same small amount $\pi/10$ toward the nearest measurement pole over the delay range $T_d\in[0,\tau_m]$, the Bloch radius $R$ becomes substantially more degraded from $R_{\rm max} = 0.64$ to a mere $R_s = 0.15$ over the same range. Such dramatic degradation of $R$ may not be mitigated by correction of the angular drift. Moreover, the Bloch radius sharply decreases linearly for even small amounts of delay, indicating that the stabilization protocol is quite sensitive to feedback delay. To be useful, this implies that the feedback delay must be a small fraction of the measurement collapse time $T_d/\tau_m \lesssim 0.2$. For typical delay times of $T_d \sim 200$ns, this mandates very slow measurements with $\tau_m \gtrsim 1\mu$s. We note that the sensitivity to delay seen here is in good qualitative agreement with a similar analysis of stabilized Rabi oscillations for a double-quantum-dot performed in Ref. \cite{Ruskov2004}.

The notable exceptions to the behavior described above are the specific target angles $\theta_T = \pm\pi/2$ on the equator of the Bloch sphere. For these angles, stabilization is excellent when $T_d = T_s = 0$, even in the non-ideal case, but rapidly becomes impossible for even small amounts of smoothing and delay. Curiously, the regions around the equator are still well-stabilized in accordance with the preceding paragraphs, but there is a sharp instability precisely at the equator. Such an instability at the equator is reminiscent of that seen in fluorescence-based qubit state stabilization protocols \cite{wangwiseman2001,Wiseman2002bayes,wang2005entfeed,Ibarcq2016}; however, it is notable that our protocol does not display such an instability until signal filtering or feedback delay are added.

\section{Trajectory Stabilization}\label{sec:trajectory}
In the preceding section we posed a stabilization problem for which the ensemble-average state is the only accessible quantity to the experimenter. That is, the experimenter has no knowledge of any detailed prior evolution of the state during the stabilization process, and only receives a fluctuating output state. 
We now consider a slightly modified posing of the problem that permits deeper investigation into the nature of the stabilization process itself.

Consider again a closed, continuous measurement feedback loop used as a storage mechanism for preparing a target qubit state on demand. However, now let the feedback loop also internally track the evolution of the qubit using the measured continuous readout. An experimenter will press a button as before, at which point the feedback loop will be terminated and the prepared qubit given to the experimenter. However, the state preparation procedure will now also report its estimate of the actual qubit state to the experimenter. The experimenter may then choose to discard the qubit and wait for a new preparation, or choose to perform an experiment immediately with the prepared state. \emph{Can the feedback parameters $\Delta_0$ and $\Delta_1$ be fixed such that a particular qubit state is reliably prepared with high probability for each single shot of the experiment?}  

Note that this trajectory-specific stabilization problem can dramatically differ from the ensemble-averaged version in the previous section, since the peak of the distribution of prepared states need not correspond to the mean of that distribution. The peak of the trajectory distribution corresponds to the most likely state to be prepared, which we can take to be the most relevant quantity. That is, the experimenter can readily discard the states far away from the peak while still keeping a substantial fraction of the prepared states. Thus, we seek to characterize the dominant distribution peak and the spread of the distribution around that peak. 

We show that for idealized Markovian feedback, perfect stabilization of individual qubit trajectories is possible in the time-continuous limit, with an identical solution for the feedback parameters as obtained for the ensemble-averaged case. We then generalize the protocol to add realistic experimental non-idealities and show that the stabilization condition may no longer be completely satisfied for individual trajectories. Nevertheless, the optimal stabilization for individual trajectories still coincides with the stabilization of ensemble averaged states possessing maximum purity, yielding again the same solution for the feedback parameters discussed in the previous section. The lack of perfect stabilization has a substantial impact on the structure of the state trajectory distribution, however, causing broadening relative to the ideal case, as well as a shift of the most probable peak away from the ensemble-average. Finally, we discuss the effect of realistic feedback delay and signal filtering on the stabilization protocol and show that it rapidly broadens the distributional width even further unless the delay time is short compared to the measurement collapse timescale.

\subsection{Stationary states for ideal Markovian feedback}\label{sec:idealtraj}
To stabilize an individual trajectory to a particular stationary point, the effect of the fluctuating noise must vanish at that point. As such, we can re-examine the Bloch Eqs.~\eqref{eq:nonidealME} in the presence of an arbitrary noise realization $\xi$ (assuming ideal limits $T_1,\,T_2 \to \infty$ and $\eta,\,R_s\to 1$ as in Sec.~\ref{sec:ideal}) and demand that $\dot{y}=\dot{z}=0$ as before. Forcing the noise term to vanish for all noise realizations $\xi$ produces the additional constraints
\begin{align}
  -y_sz_s + \tau_m\Delta_1z_s &= 0, & y_s^2 - \tau_m \Delta_1 y_s &= 0.
\end{align}
Both these constraints are automatically satisfied by the feedback parameter $\Delta_1 = y_s/\tau_m$ already derived for the ensemble-averaged case in Eq.~\eqref{eq:idealparams}. That is, at the stationary state of the ensemble average, the backaction from the noise vanishes for each individual trajectory, so each trajectory should also be independently stabilized. This result is expected because the ensemble average solution was a pure state, indicating a certainty only obtainable if noise fluctuations identically vanish. 

\begin{figure*}[t]
	\includegraphics[width=0.9\textwidth]{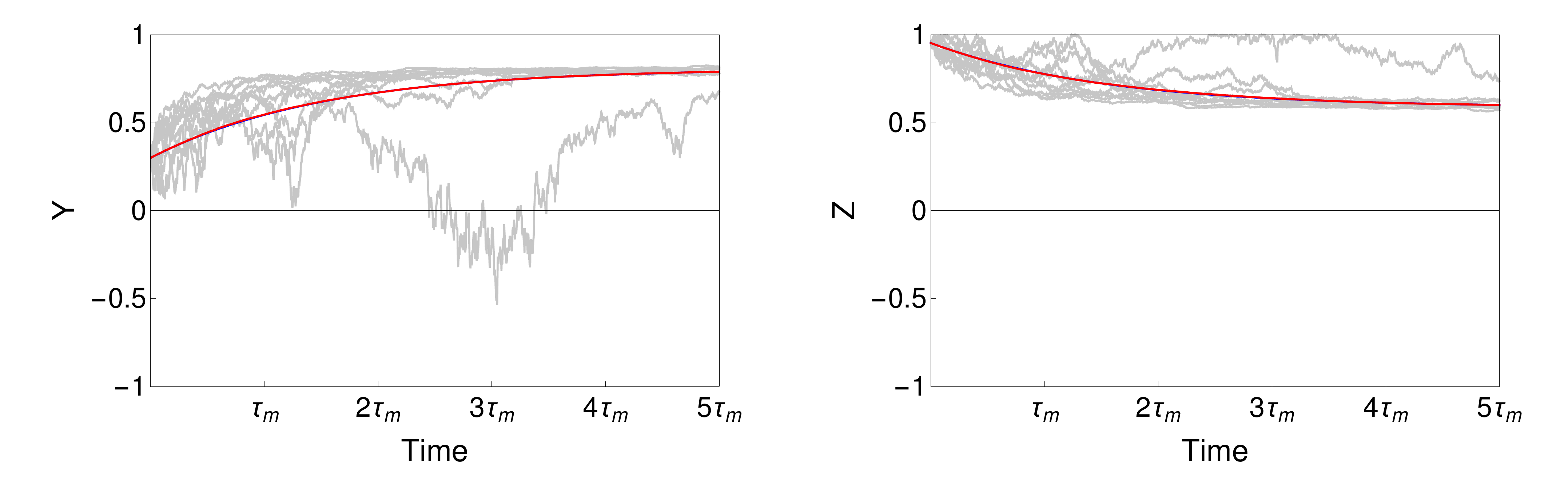}
	\caption{Ideal Markovian feedback stabilization for qubit Bloch coordinates $(y,z)=(\sin\theta,\cos\theta)$, shown (left, right), with collapse time $\tau_m = 0.2\mu$s and time step $dt = 0.5$ns. By setting the feedback parameters $\Delta_0 = -\sin(2\theta_s)/4\tau_m$ and $\Delta_1 = \sin\theta_s/\tau_m$, the initial state at $\theta=\pi/10$, $(y, z)=(0.3, 0.91)$, correctly evolves to the target stationary state with angle $\theta_s=3\pi/10$, $(y_s, z_s) = (0.81, 0.59)$, within a few measurement collapse times $\tau_m$. Analytic solutions for the ensemble-average state evolution (blue, nearly coinciding with red) are compared to the numerically simulated ensemble-average over $10^4$ trajectories (red). Individual sample trajectories (gray) roughly indicate the distributional convergence to the targeted state.}
    \label{fig:idealtraj}
\end{figure*}

\begin{figure*}[t]
	\includegraphics[width=0.9\textwidth]{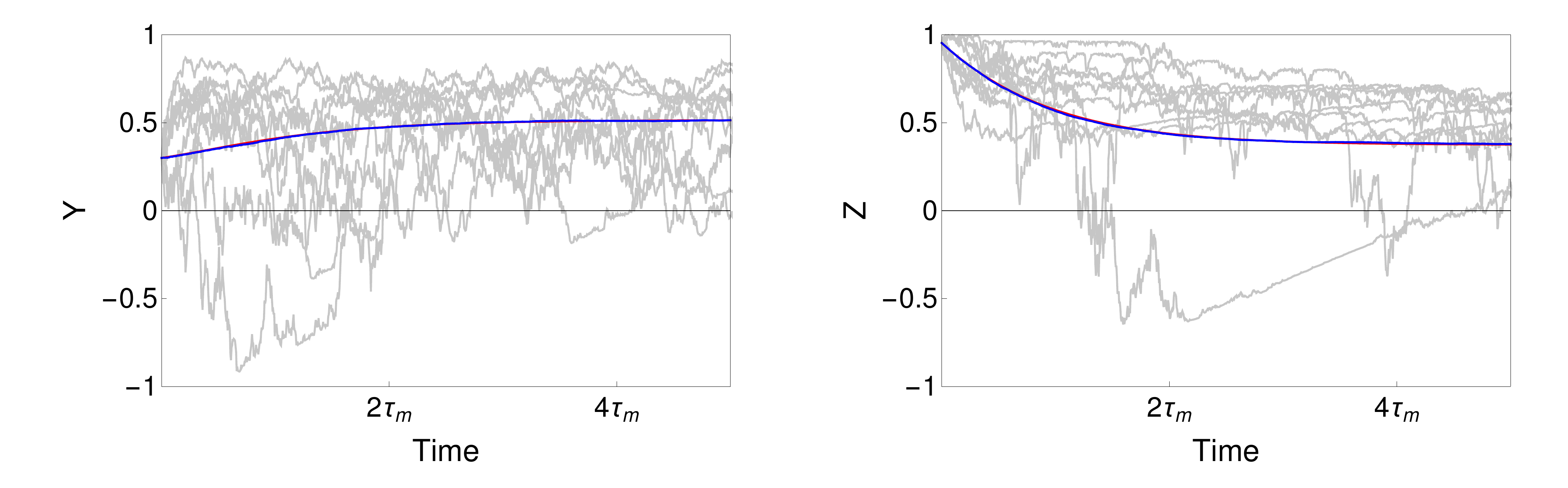}
	\caption{Non-ideal Markovian feedback stabilization for qubit Bloch coordinates $(y,z)=(R\sin\theta,R\cos\theta)$, shown (left, right), with collapse time $\tau_m = 0.2\mu$s, time step $dt = 0.5$ns, energy decay time $T_1=60\mu s$, environmental dephasing time $T_2 = 40\mu$s, and quantum efficiency $\eta=0.41$. By setting the feedback parameters $\Delta_0 = -[\sin(2\theta_s)/4\tau_mR_s^2]$ and $\Delta_1 = (\sin\theta_s/\tau_mR_s)$, with $R_s(\theta_s)$ determined by Eq.~\eqref{eq:maxR} in the text, the initial state at angle $\theta=\pi/10$ and unit radius, or $(y, z)=(0.3, 0.91)$, correctly evolves to the target stationary state with angle $\theta_s = 3\pi/10$ and radius $R_s = 0.64$, or $(y_s, z_s) = (R_s\sin\theta_s,R_s\cos\theta_s) = (0.52, 0.37)$ on average, within a few measurement collapse times $\tau_m$. Analytic solutions for the ensemble-average state evolution (blue, nearly coinciding with red) are compared to the numerically simulated ensemble-average over $10^4$ trajectories (red). Individual sample trajectories (gray) show broad fluctuation around the target state compared with the ideal case in Fig.~\ref{fig:idealtraj}.}
    \label{fig:nonidealtraj}
\end{figure*}

For time-discrete simulations, however, this ideal stability point is imperfect. The finite state jumps due to the finite time steps cause minor fluctuations around the stability point in practice, which vanish in the limit $dt\to 0$. Interestingly, near the equator with $z=0$, the width of the fluctuation distribution around the stability point with larger $dt$ narrows considerably. This improved stability at the equator is akin to balancing a rod nearly vertically by shifting its balance point laterally, which can be accomplished most efficiently when the rod is at the unstable equilibrium point at the vertical. Conversely, near the poles with $z=\pm 1$ the distribution broadens due to the destabilizing influence of the nearby pole. This converse situation is akin to balancing a rod at an extreme angle with lateral motion, which is prone to larger error since the feedback competes with the nearby stable equilibrium point. This intuition about balancing a rod also helps to explain the disastrous effect of the signal filtering and feedback delay seen at the Bloch sphere equator in Sec.~\ref{sec:delay}: If control is applied too late or in the incorrect direction for a vertically balanced rod, it will fall before the control can effectively recover. However, the stabilization of angles slightly away from the unstable equilibrium is more tolerant to delay, as the required feedback signal is already more significant.

Example trajectories are shown in Fig.~\ref{fig:idealtraj}, showing the stabilization from an initial state at angle $\theta=\pi/10$ to a target state at angle $\theta_s=3\pi/10$. Not only does the ensemble-average of $10^4$ trajectories (red) correspond to the analytical solution (blue) of the ensemble-average Markovian feedback Eq.~\eqref{eq:blochupdate}, but most individual trajectories (gray) correctly converge to the same ensemble-average stability point within a few collapse times $\tau_m = 0.2\mu$s. For simulation purposes, $dt = 0.5$ns; note that further decreasing the time step size $dt$ will only improve the convergence to the stability point, as the analytic results were derived strictly for the $dt\to 0$ limit. 

Notably, the state trajectory stability shown here behaves differently from previously considered fluorescence feedback protocols \cite{wangwiseman2001,wang2005entfeed,Ibarcq2016}, which were unable to stabilize the Bloch equator (even without signal filtering or feedback delay). The reason for the difference in the dispersive measurement case is that the measurement backaction symmetrically attracts the state toward both poles, so can balance the equator akin to the vertical balancing of a rod at an unstable equilibrium. In contrast, the fluorescence measurements considered in Refs. \cite{wangwiseman2001,Wiseman2002bayes,wang2005entfeed,Ibarcq2016} have fundamentally asymmetric backaction, which resulted in the equator becoming unstable. 

\subsection{Minimum disturbance Markovian feedback}\label{sec:nonidealtraj}
When nonidealities are added as in Sec.~\ref{sec:nonideal}, it is no longer possible to perfectly remove the backaction from the noise, even in the continuous Markovian limit $dt\to 0$. From Eq.~\eqref{eq:nonidealME} we can isolate the disturbance caused to the $y$ and $z$ coordinates per unit noise $\xi/\sqrt{\tau_m}$,
\begin{align}
  \delta y &\equiv -y_s z_s + \tau_m \Delta_1 z_s, & \delta z &\equiv (1-z_s^2) - \tau_m\Delta_1 y_s.
\end{align}
These noise-disturbances should be minimized at a stationary point $(y_s,z_s)$, and depend only on the feedback parameter $\Delta_1$. However, they do not both vanish simultaneously for any choice of $\Delta_1$ if the stabilized state is not pure. That is, 
\begin{align}\label{eq:minparams}
  \Delta_1 &\xrightarrow{\delta y = 0} \frac{y_s}{\tau_m} = \frac{R_s\sin\theta_s}{\tau_m}, \nonumber \\
  \Delta_1 &\xrightarrow{\delta z = 0} \frac{1-z_s^2}{\tau_m y_s} = \frac{1-R_s^2}{\tau_m R_s\sin\theta_s} + \frac{R_s\sin\theta_s}{\tau_m}
\end{align}
This discrepancy implies that a minimum fluctuation must persist for any choice of $\Delta_1$.

A natural resolution to this dilemma is to minimize the total noise-disturbance. That is, after defining the noise-disturbance for the entire Bloch vector $\delta \vec{s} = (0, \delta y, \delta z)$, a natural cost function is the length of this vector $C(\Delta_1) = |\delta\vec{s}|^2 = (\delta y)^2 + (\delta z)^2$. Minimizing this cost via $C'(\Delta_1)=0$ yields an optimal feedback parameter that is distinct from both single-coordinate optima in Eq.~\eqref{eq:minparams},
\begin{align}\label{eq:optparam}
  \Delta_1 &\xrightarrow{C'=0} \frac{y_s}{R_s^2\tau_m} = \frac{\sin\theta_s}{R_s\tau_m}.
\end{align}
Notably, this minimum noise-disturbance $\Delta_1$ precisely corresponds to the form found for the ensemble-average stability condition in Eq.~\eqref{eq:stationaryfeedback} when the average radius $R_s$ is made maximal via the constraint in Eq.~\eqref{eq:maxR}. We can thus understand the mysterious condition in Eq.~\eqref{eq:maxR} as physically corresponding to the requirement that the noise-disturbance be minimal for individual trajectories.

We thus completely recover the optimal results outlined for the ensemble average case in Eqs.~\eqref{eq:stationaryfeedback} and \eqref{eq:maxR}, with limit as $T_1\to\infty$ given explicitly in Eq.~\eqref{eq:nonidealparams}. This result is sensible, since the ensemble-average should be most pure when each individual trajectory is optimally stabilized at its maximum achievable purity. Pragmatically, this means that there is no difference in procedure between trying to stabilize an ensemble-averaged state with maximal purity and trying to stabilize individual trajectories to have minimum measurement disturbance in the case of Markovian feedback. The two variations of our stabilization problem produce identical parameter prescriptions for $\Delta_0$ and $\Delta_1$.

Example trajectories are shown in Fig.~\ref{fig:nonidealtraj}, showing the same stabilization task from an initial state at angle $\theta=\pi/10$ to a target state at angle $\theta_s=3\pi/10$ as in Fig.~\ref{fig:idealtraj} for comparison. Due to the nonidealities of efficiency $\eta=0.41$, energy decay time $T_1=60\mu$s, and environmental dephasing time $T_2 = 40\mu$s, the target maximum ensemble-averaged radius $R_s = 0.64$ follows from Eq.~\eqref{eq:maxR}, using the same collapse time $\tau_m = 0.2\mu$s, and time step $dt =0.5$ns. As with the ideal case, the ensemble-average of $10^4$ trajectories (red) corresponds well to the analytic solution (blue) of the ensemble-average Markovian feedback, Eq.~\eqref{eq:blochupdate}. However, the distribution of individual trajectories (gray) is now much broader around the ensemble-average stability point. Despite the chosen feedback parameters minimizing the total noise disturbance, there is still significant fluctuation within the stabilized region. 

\begin{figure}[t]
    \includegraphics[width=\columnwidth]{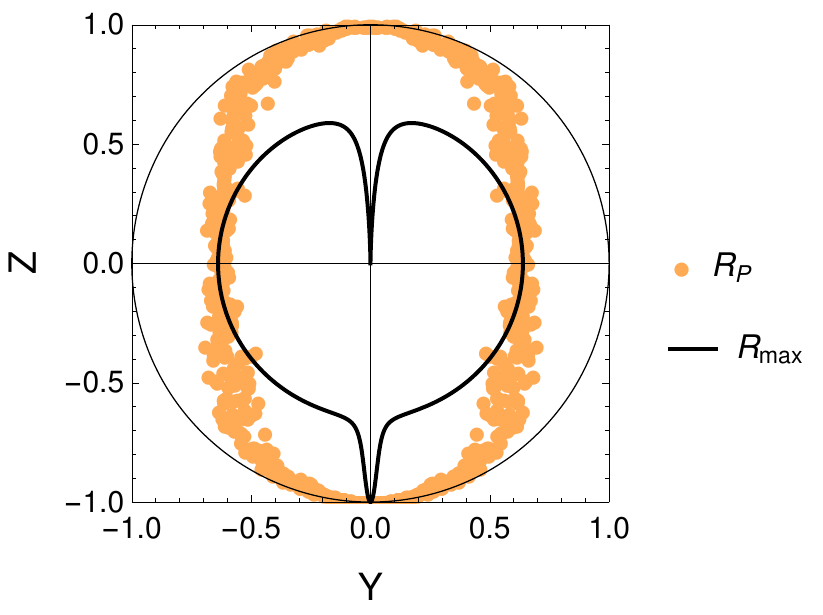}
    \caption{Comparison between targeted ensemble-average states (black curve) and most probable individual trajectories (orange dots), for parameters $T_1 = 60\mu$s, $T_2 = 40\mu$s, $\eta = 0.41$, $\tau_m = 0.2\mu$s, and $dt=10$ns. Each orange dot is the maximum histogram bin for an ensemble of $10^5$ trajectories, targeting a range of polar angles $\theta\in(0,\pi)$. The computed data has been reflected over the $y=0$ axis for visual appeal. The ensemble-averaged state (black) has constant radius of $R_E = 0.64$ except near the poles where $T_1$ effects manifest. Note that the stabilization of the excited state pole is compromised, while stabilization of the ground state pole is enhanced.  The most probable states (orange) coincide with the mean only at the equator when $z=0$ and otherwise purify near the poles, as also shown in Fig.~\ref{fig:3dhists}. }
    \label{fig:RpRe}
\end{figure}

A steady-state trajectory histogram on the Bloch plane typically shows a single dominant lobe centered near the target stability point. Notably, however, for stabilization points near the poles at $z=\pm 1$, the distribution of trajectories may bifurcate into two distinct lobes such that the ensemble-average point corresponds to the peak of neither lobe. Example histograms of cases at $\theta_s=3\pi/10$ and $\theta_s=\pi/10$ are shown in Fig.~\ref{fig:3dhists} to emphasize this splitting of stabilization lobes that can occur near a pole. For the posing of the stabilization problem in this section, an experimenter will obtain likely states at the peaks of each lobe with high probability, and will recover the ensemble-average from the posing in the last section only after averaging many of these high-probability preparations. 

Both the primary and secondary stability lobes for targeted angles near a pole tend to be more pure than stabilized states near the equator. However, averaging the lobes together preserves the same, nearly angle-independent, ensemble-average Bloch radius. As such, it is significantly more challenging to relate the peak of the dominant stabilized lobe (i.e., the most likely prepared state) to the two control parameters $\Delta_0$ and $\Delta_1$ analytically. Nevertheless, the positions of the peaks for the dominant lobes may be readily mapped out numerically by varying the ostensible target angles $\theta_s$ according to Eqs.~\eqref{eq:optparam} and \eqref{eq:nonidealparams} and computing the obtained $\theta_P$ and $R_P$ for the dominant histogram peak corresponding to those parameters. We plot the result of this numerical procedure in Fig.~\ref{fig:RpRe}, showing the purification of the most likely states near the measurement poles.

\subsection{Non-Markovian experimental considerations}\label{sec:delaytraj}

\begin{figure}[t]
    \includegraphics[width=0.7\columnwidth]{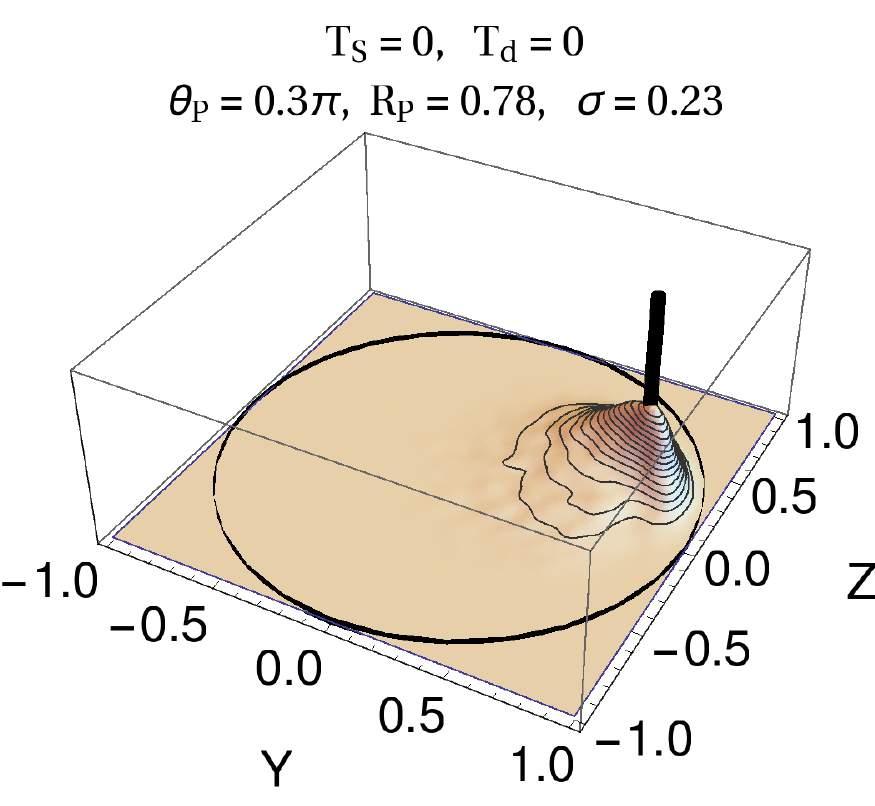}
    
    \vspace{2em}
    
    \includegraphics[width=0.7\columnwidth]{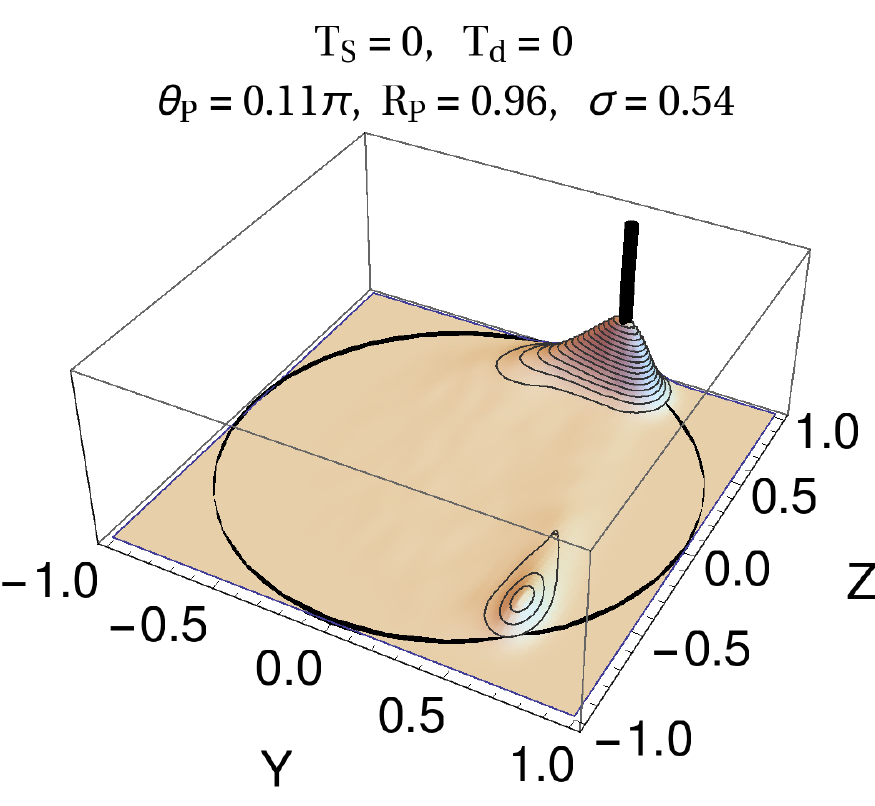}
    \caption{Nonideal trajectory histograms at steady-state in the Bloch $yz$-plane. The inefficiencies $T_1 = 60\mu$s, $T_2 = 40\mu$s, $\eta=0.41$ are included as in Fig.~\ref{fig:nonidealtraj}, with $\tau_m = 0.2\mu$s and $dt = 10$ns. Signal filtering and feedback delay are neglected here ($T_s=T_d=0$). (Top) Target ensemble-average state $\theta_s = 3\pi/10$ and $R_s = 0.64$, showing single-lobe stabilization. The actual histogram peak (black bar) is at the same target angle $\theta_P = 3\pi/10$ but with larger radius $R_P = 0.78$ than the mean, with deviation $\sigma = 0.23$ around that peak. (Bottom) Target ensemble-average state $\theta_s = \pi/10$ and $R_s = 0.64$ closer to the pole, showing double-lobe stabilization. The dominant peak (black bar) is at the shifted angle $\theta_P = 11\pi/100$ with substantially larger radius $R_P=0.96$ and deviation $\sigma=0.54$.}
    \label{fig:3dhists}
\end{figure}

\begin{figure}[t]
    \includegraphics[width=0.7\columnwidth]{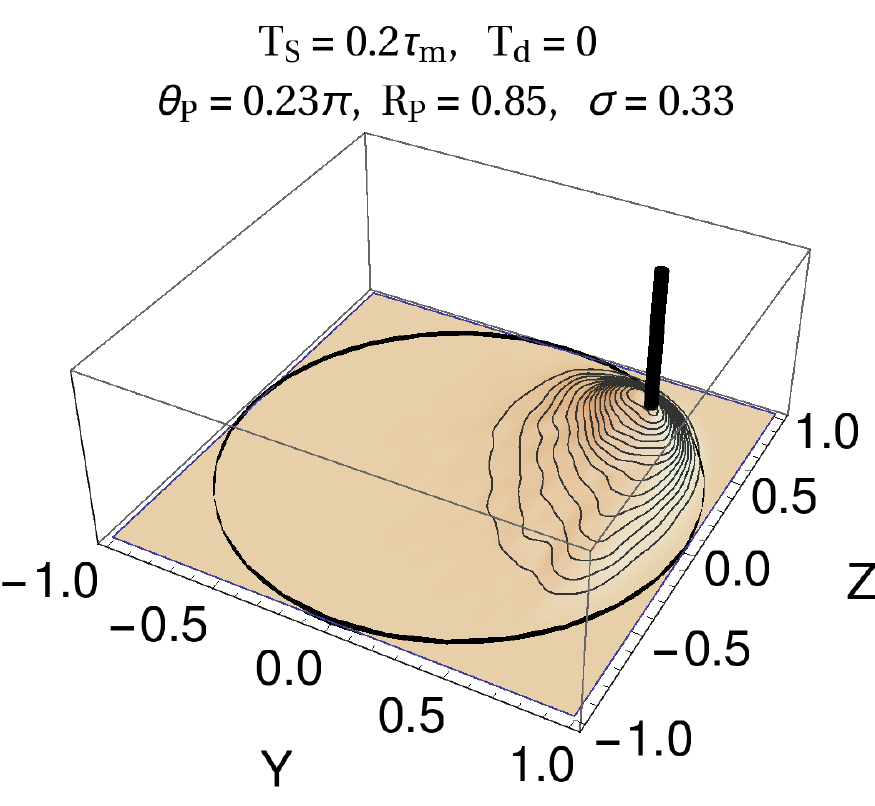}
    
    \vspace{2em}
    
    \includegraphics[width=0.7\columnwidth]{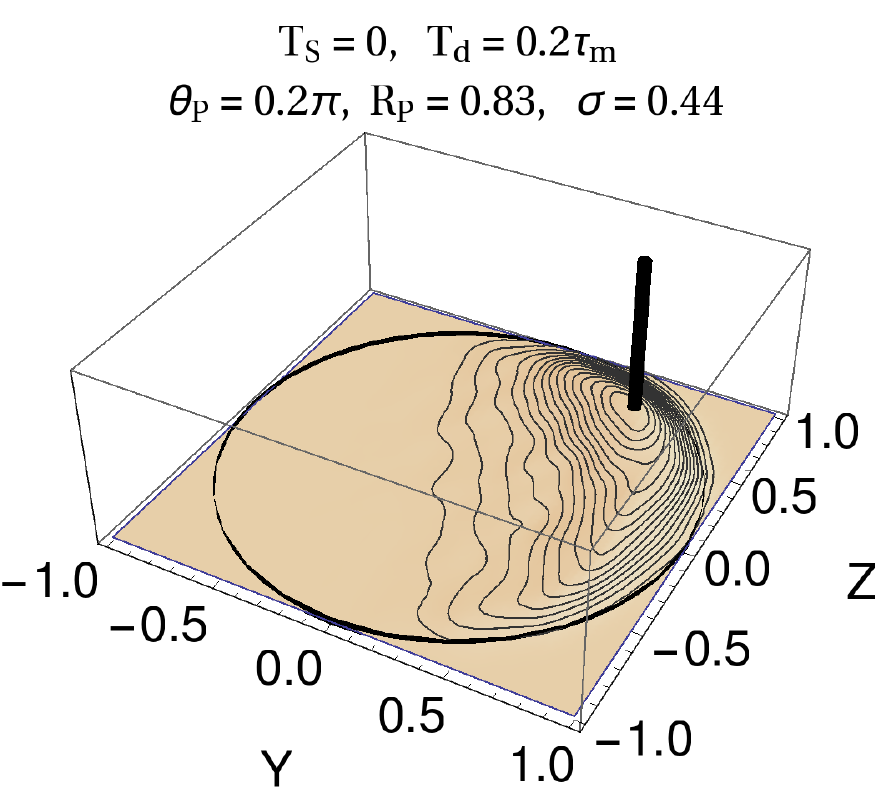}
    \caption{Nonideal trajectory histograms at steady-state in the Bloch $yz$-plane, including signal filtering and delay. The ensemble-average state $\theta_s = 3\pi/10$ and $R_s = 0.64$ is targeted using the same parameters as the top plot in Fig.~\ref{fig:3dhists} for comparison. (Top) Only single-pole low-pass signal filtering is added with exponential time constant $T_s = 0.2\tau_m$. The single lobe of Fig.~\ref{fig:3dhists} spreads into a larger region, degrading the ensemble-average purity. Most prepared states retain high purity, however, including the most probable state (black bar) with $\theta_P = 0.23\pi$ and $R_P=0.85$. (Bottom) Only feedback delay of $T_d = 0.2\tau_m$ is added. Similar to signal filtering, the single lobe spreads, but into an even larger region as indicated in Fig.~\ref{fig:delay}. The most probable state (black bar) is also similar, with $\theta_P = 0.2\pi$ and $R_P = 0.83$.}
    \label{fig:3dhistsdelay}
\end{figure}
As with the ensemble average posing in Sec.~\ref{sec:delay}, adding single-pole low-pass signal filtering and feedback delay dramatically alters the quality of the stabilization protocol. The effect can be seen most readily in the trajectory histograms, of which we show an example in Fig.~\ref{fig:3dhistsdelay} that can be directly compared with the top plot in Fig.~\ref{fig:3dhists}. The single lobe stabilization region spreads out due to both signal filtering and feedback delay into a larger region, consistent with the expectations from Fig.~\ref{fig:delay}. As such, while most prepared states will actually have relatively high purity, the ensemble average state becomes increasingly mixed. Moreover, the angular stabilization is compromised as the angular uncertainty in the broadened region becomes large. While the most probable state remains near the targeted angle, with high purity, it becomes comparable in probability to most other prepared states, and thus dubiously stabilized. 

In order to achieve the best quality stabilization, an experimenter should thus minimize the feedback delay $T_d$ relative to the measurement collapse time $\tau_m$. As estimated in Sec.~\ref{sec:delay}, for realistic delays of $T_d \sim\!200$ ns, one should use slow measurements with $\tau_m\gtrsim 1$ $\mu$s. Compared to this timescale, typical signal filtering timescales $T_s$ should also be relatively inconsequential.

\section{Conclusions}\label{sec:conclusion}

We investigated the linear feedback stabilization of a transmon qubit being continuously monitored by dispersive coupling to a microwave field. Using only a time-varying Rabi drive consisting of a constant frequency modulated by a direct linear feedback of the collected and filtered microwave signal, we showed that it is possible to prepare target qubit states on demand for any angle in the Bloch sphere. We detailed both a time-discrete and a time-continuous model for the measurement process, derived analytical solutions for the stabilization conditions, and checked the results through numerical simulations. Notably, we showed that the required feedback parameters are identical for two distinct variations of the stabilization problem: (1) on-demand preparation of a stabilized ensemble-average state, and (2) on-demand preparation of a single state trajectory with minimized total measurement disturbance. In the first situation, an experimenter does not have any information about the exact preparation shot-to-shot and so can only access the ensemble-averaged state. In the second, the experimenter also has information about the exact state that was prepared. Interestingly, in the latter the most likely states that are prepared tend to have higher purity, yet still recover the same ensemble average state as the former.

In the ideal case we showed that any target pure qubit state may be stabilized and derived the two unique feedback parameters in Eq.~\eqref{eq:idealparams} that achieve that stabilization. In the nonideal case, with experimental nonidealities of energy relaxation, environmental dephasing, and measurement inefficiency, we showed how to target ensemble average states with maximum purity by setting the feedback parameters in Eqs.~\eqref{eq:optparam} and \eqref{eq:stationaryfeedback}, which have the approximate closed form in Eq.~\eqref{eq:nonidealparams} when energy relaxation can be neglected. Expressed in Bloch coordinates, most qubit state angles may still be stabilized in the nonideal case, with a nearly angle-independent maximum radius given by Eq.~\eqref{eq:maxR}. Finally, we included the experimental nonidealities of single-pole low-pass signal filtering and feedback delay, and investigated their effect numerically. These additional effects physically arise from the feedback circuitry, and degrade the quality of the achievable state stabilization by widening the distribution of prepared state trajectories and thus shrinking the ensemble averaged state purity. Notably, feedback delay that is appreciable compared to the collapse timescale has the most damaging effect on the stabilization protocol.

This stabilization protocol is within reach of modern superconducting transmon experiments involving continuous microwave measurements. The analysis presented here should be suitable for accurately modeling a feedback loop implemented with a simple controller programmed into a field-programmable gate array (FPGA) as used in Refs. \cite{Ibarcq2013,ibarcq2015,Ibarcq2016}, or even a feedback loop consisting of a simple analog circuit as used in Refs. \cite{Korotkov2012,deLange2014}. Such a feedback stabilization experiment would prompt further investigation into the use of continuous measurement feedback for practical quantum information tasks in future work.

\begin{acknowledgments}
We thank Alexander Korotkov, Howard Wiseman, Rusko Ruskov, Joshua Combes, Shay Hacohen-Gourgy, Leigh Martin, Philippe Lewalle, and Irfan Siddiqi for helpful discussions. This work was supported by U.S. Army Research Office Grants No. W911NF-15-1-0496 and No. W911NF-13-1-0402, by National Science Foundation Grant No. DMR-1506081 including an REU extension, and by the Development and Promotion of Science and Technology Talents Project Thailand. We also acknowledge partial support by the Perimeter Institute for Theoretical Physics. Research at Perimeter Institute is supported by the Government of Canada through Industry Canada and by the Province of Ontario through the Ministry of Economic Development and Innovation.
\end{acknowledgments}

%

\appendix
\section*{Appendix}

In this appendix, we briefly include a pure state formulation of the continuous time limit for completeness. This formulation simply reproduces the conclusions already obtained from the Bloch equations in the main text. In the ideal case without experimental nonidealities, Eqs.~\eqref{eq:measop} and \eqref{eq:unop} are sufficient to describe the pure state vector update equation
\begin{align}\label{eq:pureupdate}
  \ket{\psi(t+dt)} &= \frac{\hat{U}\hat{M}_{\bar{r}}\ket{\psi(t)}}{\sqrt{\bra{\psi(t)}\hat{M}_{\bar{r}}^\dagger\hat{M}_{\bar{r}}\ket{\psi(t)}}}.
\end{align}
Performing an expansion to linear order in $dt$ of the pure state update Eq.~\eqref{eq:pureupdate} while applying the It\^o rule $dW^2 = dt$ then produces the It\^o stochastic Schr\"odinger equation
\begin{align}\label{eq:feedbacksse}
  d\ket{\psi} &= \left[i\frac{\Delta_0}{2}\hat{\sigma}_x + i\frac{\Delta_1}{4}\hat{\sigma}_x(\hat{\sigma}_z+z) + \frac{z}{4\tau_m}\hat{\sigma}_z \right. \nonumber \\
  &\qquad \left. - \frac{1+z^2+\tau_m^2\,\Delta_1^2}{8\tau_m}\right]dt\ket{\psi}, \nonumber \\
  &+ \left[i \frac{\tau_m\,\Delta_1}{2}\hat{\sigma}_x + \frac{1}{2}(\hat{\sigma}_z - z)\right]\frac{dW}{\sqrt{\tau_m}}\ket{\psi}.
\end{align}
Though this state vector representation has lower dimensionality than the density operator SME in Eq.~\eqref{eq:feedbacksme}, it reproduces the same Bloch coordinate equations in Eq.~\eqref{eq:nonidealME}, albeit with the added constraint $x^2+y^2+z^2=1$.

\end{document}